\documentclass[twocollumns]{aa} 

\usepackage{graphicx}
\usepackage{txfonts}
\usepackage{natbib}
\bibpunct{(}{)}{;}{a}{}{,} 
\usepackage{comment}

\begin{document}

   \title{Active Anomaly Detection for time-domain discoveries}

   \titlerunning{Active Anomaly Detection for time-domain discoveries}
   \authorrunning{Ishida et al.}
   
   \author{E.~E.~O.~Ishida\inst{1},
           M.~V.~Kornilov\inst{2,3},
           K.~L.~Malanchev\inst{2,3},
           M.~V.~Pruzhinskaya\inst{2},
           A.~A.~Volnova\inst{4}, 
           V.~S.~Korolev\inst{5,6}
           F.~Mondon\inst{1},
           S. Sreejith\inst{1},
           A.~A.~Malancheva\inst{7}
          \and
          S. Das\inst{8}}

   \institute{Universit\'e Clermont Auvergne, CNRS/IN2P3, LPC, F-63000 Clermont-Ferrand, France\\
        \email{emille.ishida@clermont.in2p3.fr}
         \and
        Lomonosov Moscow State University, Sternberg Astronomical Institute, Universitetsky pr.~13, Moscow, 119234, Russia\\
             \email{matwey@sai.msu.ru}
        \and
        National Research University Higher School of Economics, 21/4 Staraya Basmannaya Ulitsa, Moscow, 105066, Russia
        \and
        Space Research Institute of the Russian Academy of Sciences (IKI), 84/32 Profsoyuznaya Street, Moscow, 117997, Russia
        \and
        Central Aerohydrodynamic Institute, 1 Zhukovsky st, Zhukovsky, Moscow Region, 140180, Russia
        \and
        Moscow Institute of Physics and Technology , 9 Institutskiy per., Dolgoprudny, Moscow Region, 141701, Russia
        \and
         Cinimex, Bolshaya Tatarskaya street 35 bld. 3, Moscow, 115184, Russia
         \and
         Washington State University, Pullman, WA 99163, USA}

   \date{Received Month DD, YY; accepted Month DD, YYY}

 
  \abstract
   {}
   {We present the first evidence that adaptive learning techniques can boost the discovery of unusual objects within astronomical light curve data sets.}
   {Our method follows an active learning strategy where the learning algorithm chooses objects which can potentially improve the learner if additional information about them is provided. This new information is subsequently used to update the machine learning model, allowing its accuracy to evolve with each new information. For the case of anomaly detection, the algorithm aims to maximize the number of scientifically interesting anomalies presented to the expert by slightly modifying the weights of a traditional Isolation Forest (IF) at each iteration. In order to demonstrate the potential of such techniques, we apply the Active Anomaly Discovery (AAD) algorithm to 2 data sets: simulated light curves from the PLAsTiCC challenge and real light curves from the Open Supernova Catalog. We compare the AAD results to those of a static IF. For both methods, we performed a detailed analysis for all  objects with the $\sim$2\% highest anomaly scores.}
   {We show that, in the real data scenario, AAD was able to identify $\sim$80\% more true anomalies than the IF. This result is the first evidence that AAD algorithms can play a central role in the search for new physics in the era of large scale sky surveys.}
   {}


   \maketitle
%

\section{Introduction}

The detection of new astronomical sources is one of the most anticipated outcomes from the next generation of large scale sky surveys. Experiments like the Vera Rubin Observatory Legacy Survey of Space and Time\footnote{\url{https://www.lsst.org/}} (LSST) are expected to continuously monitor large areas of the sky with remarkable deliberation, certainly leading to the detection of unforeseen astrophysical phenomena. At the same time, the volume of data gathered every night will also increase to unprecedented levels, rendering serendipitous discoveries unlikely. In the era of big data, most detected sources will never be visually inspected, and the use of automated algorithms is unavoidable. 

The task of automatically identifying peculiar objects within a large set of normal instances has been highly explored in many areas of research \citep{aggarwal2016outlier}. This has led to the development of a number of machine learning (ML) algorithms for anomaly detection (AD) with a large range of applications \citep{Mehrotra:2017:ADP:3207800}.  In astronomy, these techniques have been largely applied to areas like the identification of anomalous galaxy spectra \citep{2017MNRAS.465.4530B}, problematic objects in photometric redshift estimation tasks \citep{2015MNRAS.452.4183H}, light curves (LCs) of transients \citep{Zhang_2018, 2019arXiv190511516P} and variable stars \citep[e.g.,][]{2009arXiv0905.3428R, Nun2014, Giles2019}, among others. 

Despite encouraging results, the application of traditional AD algorithms to astronomical data scenarios is far from trivial. Most of these strategies involve constructing a statistical model for the nominal data and identifying  objects which significantly deviate  from this model  as anomalous. Once identified, these sources are subjected to further scrutiny by an expert who confirms (or not) the discovery of a new phenomenon. However, frequently, a statistical anomaly may be the result of observational defects or other spurious interference which are not scientifically interesting,  leading to a high rate of candidates which turn out to be well known nature,  despite their high anomaly scores. This wrong identification results in a proportional fraction of resources, and research time, spent to further investigate these non-peculiar objects.

Since measuring the details of a new source often requires the allocation of spectroscopic follow-up resources, the development of anomaly detection strategies able to deliver a low rate of objects from scientifically well known categories is an exceedingly important task. This task will be made more crucial in the light of the upcoming generation of telescopes, which will drastically increase the volume of nominal data and, in the process, pose a challenging anomaly detection task. In ML jargon, this would require an adaptive recommendation system which is able to optimally exploit a given ML model by carefully choosing objects which can significantly influence the results, if more information about them is provided.

Active learning (AL) is a subclass of ML algorithms designed to guide such an optimal allocation of labelling resources in situations where labels are expensive and/or time consuming \citep{ALbook}. 
It has been widely applied in many real world situations and research fields,  e.g. natural language processing \citep{thompson1999}, spam classification \citep{debarr2009}, cancer detection \citep{liu2004} and sentiment analysis \citep{kranjc2015}. In the context of large scale photometric surveys, this translates into a recommendation system for planning on the distribution of follow-up resources --- given a particular scientific goal. Prototypes using this underlying philosophy for supervised learning tasks were applied to the determination of stellar population parameters \citep{solorio2005},  supervised classification of variable stars \citep{richards2012b}, microlensing events \citep{xia2016}, photometric redshift estimation \citep{vilalta17}, supernova photometric classification \citep{ishida2019} and determination of galaxy morphology \citep{Walmsley2019}.

In this work, we present the first application of AL for AD in astronomical data. Similar strategies have already been reported, with encouraging results, in the identification of anomalous behaviour dangerous to web services \citep{Fan2012}, intrusion identification in cloud systems \citep{Ibrahim2019} and detection of anomalous features in building construction~\citep{Wu2019} --- to cite a few. Despite this successful track record, the particular characteristics of astronomical data, more specifically that of astronomical transients (errors in measurements, influence of observation conditions, sparse, non-periodic  and non-homogeneous time-series, etc.) makes this demonstration an important milestone in the exploitation of such techniques by the astronomical community. As a proof of concept, we applied the active anomaly detection (AAD) strategy, proposed by~\citet{Das2017} to two different data sets: simulated light curves from the Photometric LSST Astronomical Time-series Classification Challenge\footnote{\url{https://www.kaggle.com/c/PLAsTiCC-2018}} \citep[PLAsTiCC, ][]{PLAsTiCCdata} and real light curves from the Open Supernova Catalog\footnote{\url{https://sne.space/} (OSC;~\citealt{2017ApJ...835...64G}).} Used in combination with a traditional isolation forest (IF)  algorithm, the method allows increasingly larger incidence of true positives among objects presented to the expert --- enabling a better allocation of resources with the evolution of a given survey. 

In what follows, we present the data and the pre-processing analysis in Section~\ref{sec:data}. Section~\ref{sec:method} describes the AAD algorithm and its implementation, and the results are presented in Section~\ref{sec:results}. Finally, we present our conclusions and discuss implications for future large scale astronomical surveys in Section~\ref{sec:conclusions}.


\section{Data}
\label{sec:data}

This work focuses on finding anomalies within transient light curves data sets. Our experiments were performed in simulated as well as real data sets.

Our real data sample comes from the OSC. This is a public repository containing supernova (SN) light curves, spectra and metadata from a range of sources. It is also known to contain some percentage of non-supernovae contaminants ~\citep{2017ApJ...835...64G, 2019arXiv190511516P}  --- which makes it well suited for our purposes. 

The current analysis is based on the data set\footnote{Data and pre-processing pipeline for OSC are available at \url{http://snad.space/osc/}.} first presented in~\citet{2019arXiv190511516P}. Therefore, the detailed description of quality cuts, data selection process, and pre-processing pipeline are given there. For clarity we describe the main steps of the data preparation below.

From the OSC catalogue, we extracted objects with LCs in $BRI$~\citep{1990PASP..102.1181B}, $g'r'i'$ or $gri$ filters. We assumed that $g'r'i'$ filters are very similar to $gri$ and that the coefficients of their transformation equations are quite small~\citep{1996AJ....111.1748F,2006AN....327..821T,2007ASPC..364...91S}. Light curves originally observed in $BRI$ filters were converted to $gri$ using Lupton’s transformation equations\footnote{\url{http://www.sdss3.org/dr8/algorithms/sdssUBVRITransform.php}}.

The simulated data used in this work is a sub-sample of the light curves prepared for the PLAsTiCC data challenge, which was constructed to mimic the data scenario that we will encounter after 3 years of LSST observations. In order to build a data environment similar to the one we found in the OSC, we restricted our sample to 6 classes (SN Ia, SN II, SN Ibc, SN Ia-91bg, binary microlensing and pair-instability SN (PISN))\footnote{ A detailed description of the astrophysical models is given in \citet{PLAsTiCCmodels}.}. The entire PLAsTiCC test set was subjected to the light curve fitting procedure described in Section \ref{subsec:GP}.

\subsection{Light curve fit}
\label{subsec:GP}

In order to obtain a homogeneous input data matrix for the ML algorithms, all LCs were submitted to a \textsc{Multivariate Gaussian Process\footnote{\url{http://gp.snad.space}} pipeline. Instead of approximating the light curves in different filters independently, \textsc{Multivariate Gaussian Process} takes into account the correlation between different bands, approximating the data by a Gaussian process (GP) in all filters with one global fit. The kernel used in our implementation is composed of three radial-basis functions,
\begin{equation*}
    k_i(t_1, t_2) = \exp{\left(-\frac{(t_2-t_1)^2}{2\,l_i^2}\right),}
\end{equation*}
where $i$ denotes the photometric band, and $l_i$ are the parameters of Gaussian process to be found from the light curve approximation. In addition, \textsc{Multivariate Gaussian Process} includes 6 constants, three of which are unit variances of the basis processes and the other three describe their pairwise correlations. In total, the \textsc{Multivariate Gaussian Process} has 9 parameters to be fitted.}

The approximation procedure was done in  flux space. For each object, we only took those epochs that lie within the interval $[-240, +240]$ days since maximum in $r$ band, averaging measurements within a 1-day time-bin.

Each object was characterized by 374 features. The feature set included 10 parameters of the \textsc{Multivariate Gaussian Process} (9 fitted parameters of the kernel and the final log-likelihood),  the LC maximum flux and normalized GP results within $[-20, +100]$ days since maximum brightness in $r$-band in steps of 1 day, concatenated according to their effective wavelength\footnote{All the feature extraction scenarios reported in \citet{2019arXiv190511516P} were tested. The one described here produced the most significant improvements when compared to the static isolation forest algorithm.}.

After applying these steps to the OSC, we visually inspected the results and eliminated bad fits, obtaining a final set of 1999 objects\footnote{The quality cuts described in \citet{2019arXiv190511516P} aim to ensure the best behaviour of the Gaussian process regression. In future surveys such as LSST, with better quality and homogeneity of data, these can certainly be made less strict.}. 

For the PLAsTiCC data, we automatically removed all objects for which Gaussian Processes fit was unsuccessful, i.e. likelihood maximization procedure was unable to converge. A total of 7223 objects survived this  pre-processing pipeline.


\section{Methodology}
\label{sec:method}

In order to compare the AAD results with those obtained with a traditional AD method, and with a blind search, we performed a detailed analysis of all instances within $\sim$2\% highest anomaly scores (145 objects for the simulated and 40 objects for the real data). In the simulated data, this process was automatic. Once we selected the classes which represented anomalies (see Section \ref{sec:results}) the algorithm was able to read the labels directly from the data file. For the OSC data, we recruited a team of 2 human experts, with extensive experience in observational and theoretical aspects of SN science, to carefully analyse each one of the 40 candidates. These specialists performed a thorough  investigation of each candidate -- including consultation of external literature -- and were not involved in the development or implementation of the AAD strategy.

Anomaly scores were obtained according to 3 different strategies: random sampling (RS), IF (Section \ref{subsec:IsoForest}) and AAD (Section \ref{subsec:AAD}). The screening described above allowed us to coherently estimate the rate of  scientifically interesting candidates for all these strategies. Each candidate was considered anomalous or nominal according to the guidelines described in Section \ref{subsec:def_anom}.

In essence, we follow a methodological strategy similar to the one used in internet search engines, where the relevance of a document is judged with respect to the information the user needs, i.e., the capability of solving the user's real-world problem, not just the presence of the queried words \citep[see e.eg., ][]{Manning2008}). Similarly, when evaluating different algorithms, we start from a statistically identified anomaly candidate but leave the final judgement to the experts -- allowing the system to learn the connection between the data and the user specific interests.

\subsection{Isolation forest}
\label{subsec:IsoForest}

Anomalies are identified as patterns or individual objects within a data set which do not conform to a well defined notion of ``normal''   \citep{chandola2009}. Starting from this definition, popular AD techniques begin by modelling the nominal data (defining what is normal) and subsequently identifying anomalies as samples which are unlikely to be generated by the determined model. In real data problems, this task is non-trivial, since the underlying statistical distribution guiding the data generation process can be quiet complex. It is possible to avoid the need for modelling the nominal data by using distance-based techniques. In this paradigm one starts with the hypothesis that anomalous instances are likely to be far from the normal ones in the input feature space. Thus, by calculating the distance between every possible pair of objects in the data set, it is possible to select samples which, on average, are farther from the bulk of the data set. Such a strategy avoids the need for defining a complete statistical model for the normal data but can still be computationally very expensive for large data volumes \citep{ADreview}.

IF is a tree-based ensemble\footnote{Ensemble methods are those that use a collection of learners in a synergistic manner in the formulation of the final prediction.} method first proposed by \citet{liu2008isolation}. It was inspired by distance-based techniques, and thus considers anomalies as data instances which are isolated from the bulk of the data set in a given feature space. However, this isolation is determined locally by training a randomized decision tree \citep{RFthesis}. In a sequence of steps, the algorithm randomly selects a subset of the data, input features, and split points (decision boundaries or nodes). The feature space is then sequentially subdivided into cells, with the number of sequential cuts determining the path length from the initially large feature space (root) to each final cell (leaf or external node). 
In this context, anomalies are identified as objects with the smallest path length between the root and an external node. In other words, anomalies are identified as objects which become isolated in a cell more quickly. The combination of results from a number of trees built with different sub-samples makes it robust to overfitting. By exploiting the fact that anomalies are, by definition, rare and prone to isolation, the method avoids the need for expensive distance calculations or statistical modelling of the normal instances.

\begin{figure*}
    \centering
    \begin{minipage}{0.475\textwidth}
    \includegraphics[width=\textwidth]{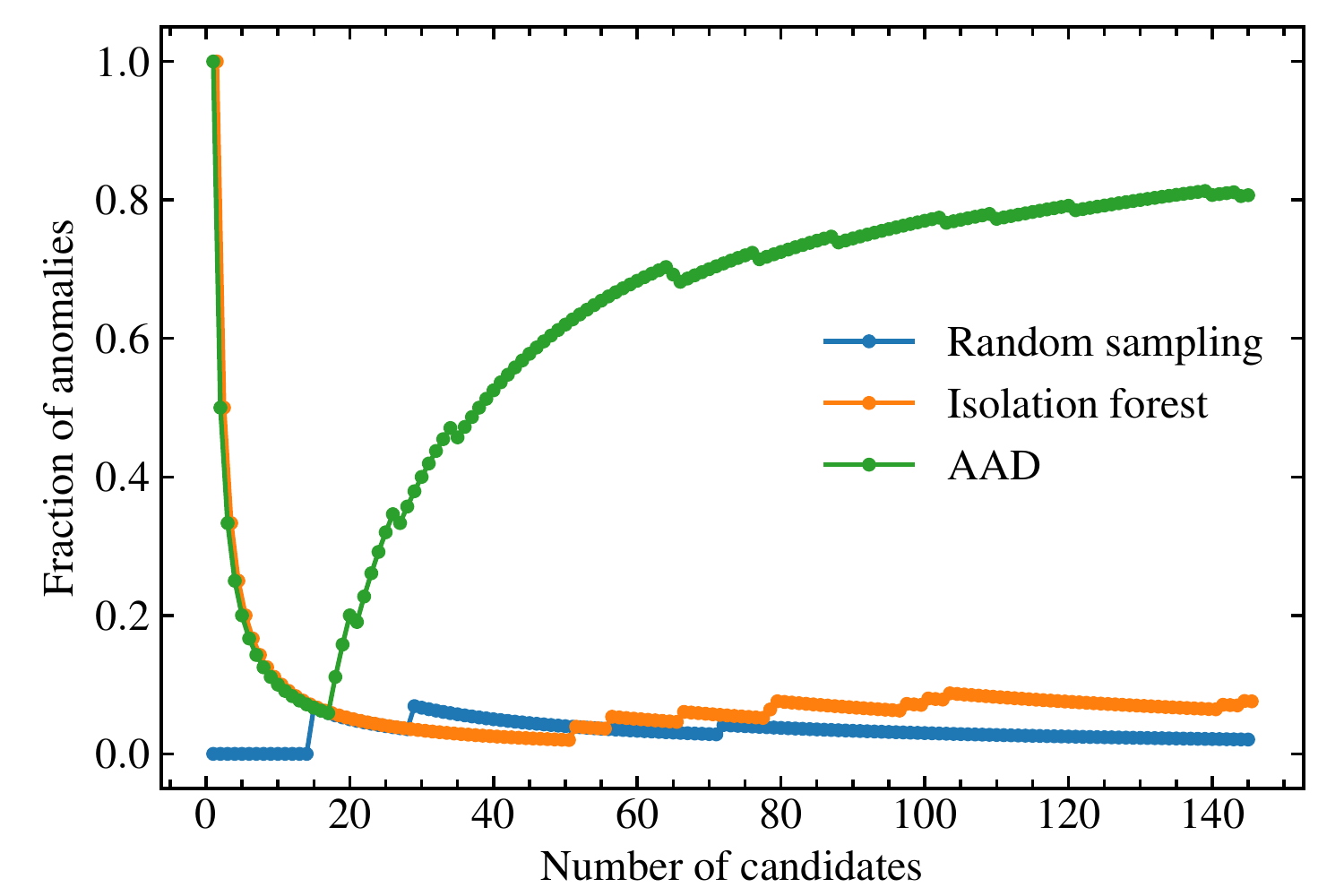}
    \end{minipage}
    \begin{minipage}{0.475\textwidth}
    \includegraphics[width=\textwidth]{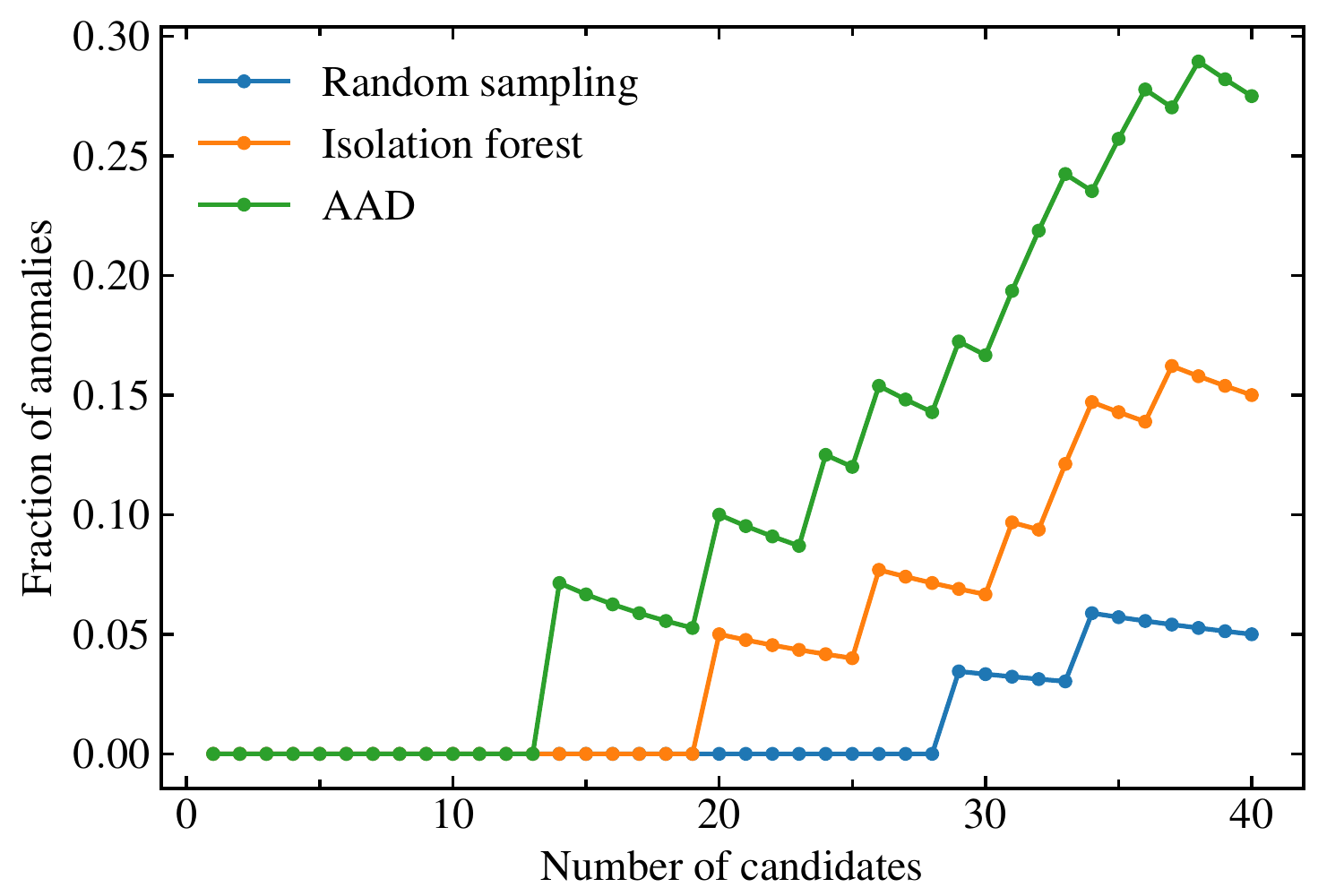}
    \end{minipage}
    \caption{Fraction of anomalies as a function the total number of  candidates. The plot shows results obtained with random sampling (blue), isolation forest (orange) and active anomaly discovery (green) algorithms. \textbf{Left:} results from the simulated PLAsTiCC data set. \textbf{Right:} results from the real OSC data.}
    \label{fig:eval}
\end{figure*}

\subsection{Active Anomaly Detection}
\label{subsec:AAD}

AL algorithms allow expert feedback to be incorporated into the learning model in an iterative manner and, consequently, improve the accuracy of the predicted results. As such, they work in conjunction with a traditional ML strategy which must either be sensitive to small changes in input information or allow the incorporation of such knowledge in subsequent fine tuning of the model. Decision trees fulfill these requirements \citep[see e.g.,][]{Loh2014}. Moreover, for the specific case of AL for AD tasks, ensemble methods are especially significant.

Ensemble methods for AD rely upon the assumption that anomalies will have a higher anomaly score across the entire ensemble, while nominal samples will be assigned lower ones --- despite  values of the scores themselves being different among ensemble members. This allows us to define a weight vector $\pmb{w}$, whose elements denote the impact of different members of the ensemble in the final anomaly score. In the case of $N$ members with perfect predictions, this will be a uniform vector, $w_i = 1/\sqrt{N}$ for $i \in [1,N]$. In a more realistic scenario, certain members will be better predictors than others and we can translate this behaviour by assigning larger weights to more accurate predictors and lower ones to noisier members of the ensemble (see Figure 1 of \citealt{Das2018}).

Active Anomaly Discovery\footnote{\url{https://github.com/shubhomoydas/ad_examples}} \citep[an AAD algorithm proposed by][]{Das2017} exploits this adaptability in order to fine tune the ensemble according to a specific definition of anomaly, as pointed out by the expert through a series of labeled examples. The algorithm starts by training a traditional IF and then presents  the candidate with the highest anomaly score to a human annotator for classification. If the expert judges the candidate to be an anomaly, the state of the model does not change and the next highest scored candidate is presented. Whenever a given candidate is flagged as nominal, the model is updated by re-scaling the contribution of each leaf node (changes in $\pmb{w}$)  to the final anomaly score. This slight modification preserves the structure of the original forest while adapting the weights to ensure that labelled anomalies are assigned higher anomaly scores than labelled nominal instances. Further details about the algorithm are given in Appendix \ref{ap:AAD} and in \citet{Das2017, Das2018}.

\begin{figure*}
\begin{center}
\includegraphics[trim={0 11cm 0 0},clip,scale=0.65]{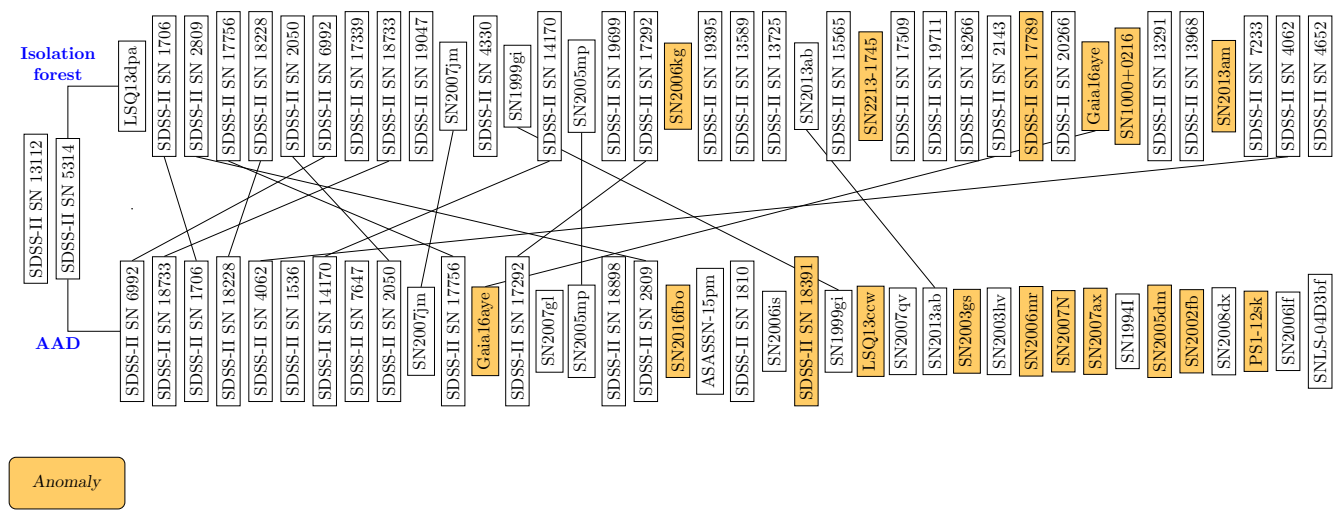}
\caption{Comparison between the outputs of isolation forest and active anomaly discovery (AAD) algorithms when applied to the Open Supernova Catalog data. Rectangles contain object names selected candidates in the order of their importance. The yellow boxes show anomalies that were visually confirmed. Solid lines indicate the objects in common for both branches.}
\label{workflow}
\end{center}
\end{figure*}

\subsection{Defining anomalies}
\label{subsec:def_anom}

The definition of anomaly strongly depends upon the goals and objectives of the researcher. In this work, we are mainly interested in identifying non-SNe contamination and/or SNe with unusual properties \citep{Mili2018}. Non-SNe objects can be divided into cases of mis-classification (quasars, binary microlensing events, novae, etc.) or  completely new classes of objects. 
We did not considered as anomalies cases of possible miss-classifications  due to signals which were too weak to allow a confident conclusion regarding the nature of the transient. These cases cannot be carefully studied due to low signal-to-noise ratio and, therefore, are not astrophysically interesting.   

We consider as unusual SNe, those objects that were proved to be peculiar by previous studies. These could be any kind of peculiarities: a signature of interaction with the circumstellar medium (CSM), an unusual rise or decline in the light curve rate or any other features which are not representative of the corresponding SN type.

 The anomalous cases included in our simulated data were chosen to represent different classes of anomalies: SNe Ia-91bg as an  example of a rare type of SN (47 objects), binary microlensing events as examples of mis-classifications (45 objects) and pair-instability SNe as a representative of "new physics" (184 objects). In summary, the simulated data contains $\sim 4\%$ (275) anomalies and $\sim 96\%$ (6958) nominal objects\footnote{We emphasise that we cannot calculate such percentages for the OSC data, since it would require our experts to perform a detailed analysis of all 1999 objects.}.

For data from the OSC, we consider as anomalous super-luminous SNe and SNe of rare types. Super-luminous SNe (SLSNe,~\citealt{2012Sci...337..927G}) have an absolute peak magnitude $M < -21$ mag, which is 10--100 times brighter than standard SNe. They are sometimes divided into three broad classes: SLSN-I without hydrogen in their spectra, hydrogen-rich SLSN-II that often show signs of interaction with CSM, and finally, SLSN-R, a rare class of hydrogen-poor events with slowly evolving LCs, powered by the radioactive decay of $\rm ^{56}Ni$. Due to its anomalous luminosity, SLSNe are becoming  important probes of massive star formation in the high-redshift Universe and may important cosmological probes, similar to Type Ia SNe~\citep{2014ApJ...796...87I} --- although only a couple of dozen events have been observed so far~\citep{2018SSRv..214...59M}. The physics that drives this diverse class of SNe is not clearly understood, making it paramount to increase the number of   observations.

As examples of SNe types, we considered: Ibn~\citep{2008MNRAS.389..113P}, II-pec~\citep{1991A&A...243..155L}, broad-lined Ic SNe associated with gamma-ray bursts~\citep{2017AdAst2017E...5C} and low-luminosity IIP SNe~\citep{2018MNRAS.473.3863L}. We also add to this category 91T-like, 91bg-like Type Ia  and extreme thermonuclear supernovae (e.g. Type Iax supernovae;~\citealt{2013ApJ...767...57F, 2017hsn..book..317T}). Type 1991bg SNe are characterized by a red colour at maximum light and low luminosities. Type 1991T SNe, on the other hand, shows a slow decline after maximum light and high-peak luminosities. The contamination due to the presence of 1991bg-like and 1991T-like SNe in cosmological samples can affect the measurements of dark energy parameters. This is extremely important for large surveys like the LSST, which aims to constrain cosmological parameters using the bulk of normal Type Ia SNe.
All non-physical effects, e.g. artefacts of interpolation, were not considered as anomalies.

The above criteria were designed to serve as an example of the kind of requirements one might impose to the AAD algorithm. These will certainly vary depending on the research goal, available labelling resources and the data at hand. However, for the purposes of this work, the exact anomaly definition serves merely to illustrate the flexibility of our framework. The global behaviour of exercises using different anomaly criteria should resemble those presented in Section~\ref{sec:results}.


\section{Results}
\label{sec:results}

We first report results from applying our method to the subset of the PLAsTiCC data described in Section \ref{sec:data}. Figure \ref{fig:eval} (left panel) shows the percentage of identified anomalies as a function of proposed candidates. This figure was created considering objects in decreasing order of anomaly scores (for IF) and following the order in which they were presented as candidates (for AAD and RS). Considering a total of 145 candidates ($\sim$2\% of the entire data set), RS found 3 PISNe ($\sim$2\%), IF detected 7 binary microlensing events and 4 PISNe ($\sim$8\%) among the objects with highest anomaly score, while AAD was able to indicate as anomalous, 5 binary microlensing events and 112 PISNe ($\sim$81\%). Moreover, AAD was also able to identify interesting objects much sooner than IF. The first object found by both methods was a binary microlensing event. The second true anomaly appears at 51$^{\rm st}$ place among the IF candidates but at $19^{\rm th}$ place among the AAD suggestions. Considering that in the real case, the analysis of each anomaly candidate would require the use of expensive spectroscopic telescope time, these results demonstrate how AAD can be a valuable tool in the allocation of such resources.

In order to demonstrate the flexibility of the AAD algorithm to adapt to the anomaly definition set by the expert, as stated in Section \ref{sec:method}, we also ran the AAD algorithm with a different anomaly definition. In the case where the expert would flag only binary microlensing events as anomalous, the AAD algorithm returned 11 true positives (in comparison with 5 found using the more broad anomaly definition) --  more than doubling the success rate of a very narrow search. This confirms that the method is able to adapt to the type of anomaly which is interesting to the expert and increase the fraction of candidates which might be worth the effort of further investigation.

Despite the encouraging results obtained from the simulated data, we must emphasise that the analysis of real data presents a much more complex scenario. In order to confirm if the AAD performance holds when dealing with real observations, we performed the same analysis in data from the OSC. Results are presented in the right panel of  Figure \ref{fig:eval}. In this scenario, 2\% of the entire data set corresponded to $\sim$40 objects. RS achieved a maximum anomaly detection rate of $\sim$5\% (2 objects). IF was able to boost this to $\sim$15\% (5 objects) while AAD identified $\sim$27\% of true positives (11 objects). This represents an increase of $\sim 80\%$ in the number of true anomalies detected for the same amount of resources spent in scrutinising candidates. Moreover, similar to what we found in the simulated data, although both strategies require a \textit{burn-in} period to start identifying interesting sources,  AAD presents the first anomaly much earlier ($14^{\rm th}$) place, in comparison to $20^{\rm th}$ place for IF).  The full list of identified anomalies is provided in Table~\ref{tab:outliers}, and a sub-set of their light curves is presented in Appendix~\ref{ap:figs}. 

A more detailed comparison between IF and AAD results is displayed in Figure \ref{workflow}. The diagram shows the identification of  candidates presented to the expert by IF (top) and AAD (bottom). The first two objects are the same for both algorithms, with a discrepancy starting only from the third one. Candidates are ordered by their scores for IF, from left to right. For AAD, they correspond to the highest anomaly score for successive iterations of the AL loop. Anomalies confirmed by the experts are highlighted in yellow. The plot clearly illustrates not only the higher incidence of anomalies for AAD vs IF (11 vs 6), but also the larger density among latter candidates. The lines connecting objects that are present in both branches show that the first half of the list contains many objects in common between the two algorithms. On the other hand, the second half of the AAD list contains anomalies which are absent in the upper branch. This demonstrates that the algorithm is able to adapt to the definition of anomaly according to the feedback received from the expert also in a real data scenario. Moreover, one of the most obvious peculiar objects in our sample is a binary microlensing event, Gaia16aye. It was assigned the  $33^{\rm rd}$ highest anomaly score by isolation forest, while being the first real anomaly presented by AAD --- in the $14^{th}$ iteration. These results provide the first evidence that adaptive learning algorithms can be important tools in planning optimised distribution of resources in the search for peculiar astronomical objects. 


\section{Conclusions}
\label{sec:conclusions}

The next generation of large scale sky surveys will certainly detect a variety of new astrophysical sources. However, since every photometrically observed candidate requires further investigation via spectroscopy, the development of automated anomaly detection algorithms with low incidence of false positives is crucial. Moreover, such algorithms must be able to detect scientifically interesting anomalies --- as opposed to spurious features due to observing conditions or errors in the data processing pipeline. Active learning methods are known to perform well in such data scenarios. They represent a class of adaptive learning strategies where expert feedback is sequentially incorporated into the machine learning model, allowing high accuracy in prediction while maintaining the distribution of analysis resources under control.

We report results supporting the use of active learning algorithms in the allocation of resources for astronomical discovery. We use simulated and real light curves as benchmarks to compare the rate of true anomalies discovered by a traditional isolation forest algorithm to those identified by active anomaly detection \citep{Das2017}. 

We show that active anomaly detection is able to increase the incidence of true anomalies in real data  by $80\%$ when compared to static isolation forest. Moreover, the algorithm can adapt to the definition of anomaly imposed by the expert --- which leads to a higher density of true positives in later iterations. This ensures not only a larger number of peculiar objects in total, but also guarantees that each new scrutinised source will, in the long run, contribute to the improvement of the learning model. In this context, not even the resources spent in analysing false positives, in the beginning of the survey, are wasted. 

In order to ensure a reliable estimation of true positive rates, we presented a controlled real data  scenario in the form of a catalogue containing 1999 fully observed supernova light curves. 
This allowed visual confirmation of all the objects within $2\%$ highest anomaly scores for all the algorithms. As an example of the potential which active learning techniques can unravel from legacy data, we highlight that the discovery of an important astrophysical contaminant (the binary microlensing event Gaia16aye) was presented to the expert much earlier following the active strategy when compared to its static counterpart (14$^{\rm th}$ vs 33$^{\rm rd}$ highest anomaly score). Moreover, results from simulated data confirmed that the algorithm is flexible enough to allow the adaptation of the anomaly definition according to the interest of the expert -- something which is not possible within the traditional anomaly detection paradigm. We acknowledge that important issues  need to be further addressed (e.g. the variability of results for different feature extraction methods, stream mode learning and scalability). Nevertheless, results presented here support the hypothesis that adaptive techniques can play important roles in the future of astronomy.

\section*{Acknowledgements}

E.~E.~O.~Ishida and S.~Sreejith acknowledge support from CNRS 2017 MOMENTUM grant under project \textit{Active Learning for Large Scale Sky Surveys}. M.~Pruzhinskaya and M.~Kornilov are supported by RFBR grant according to the research project 18-32-00426 for anomaly analysis and LCs approximation.  K.~Malanchev is supported by RBFR grant 20-02-00779 for preparing the Open Supernova Catalog and PLAsTiCC data. A.~Volnova acknowledges support from RSF grant 18-12-00522 for analysis of interpolated LCs. We used the equipment funded by the Lomonosov Moscow State University Program of Development. The authors acknowledge the support from the Program of Development of M.V. Lomonosov Moscow State University (Leading Scientific School "Physics of stars, relativistic objects and galaxies"). This research has made use of NASA's Astrophysics Data System Bibliographic Services and following {\sc Python} software packages: {\sc NumPy}~\citep{numpy}, {\sc Matplotlib}~\citep{matplotlib}, {\sc SciPy}~\citep{scipy}, {\sc pandas}~\citep{pandas}, and {\sc scikit-learn}~\citep{scikit-learn}.

\onecolumn
\begin{longtable}{lccp{2.2cm}cl}
\caption{Anomalies identified by IF and AAD algorithms.}
\label{tab:outliers}\\
\hline
SN name &$\alpha$ &$\delta$ &Type &z &References  \\
\hline 
\multicolumn{6}{c}{Isolation forest} \\ 
\hline
SN2006kg         & 01:04:16.98 & $+$00:46:08.9 & AGN & 0.230 & \cite{2006CBET..688....1B,2011AA...526A..28O}\\
SN2213-1745      & 22:13:39.97 & $-$17:45:24.5 & SLSN-R & 2.046 & \cite{2012Natur.491..228C}\\
SDSS-II SN 17789 & 01:29:16.13 & $+$00:42:37.9 & SLSN & & \cite{Sako2018}\\
Gaia16aye        & 19:40:01.10 & $+$30:07:53.4 & ULENS, CV & & \cite{2016ATel.9376....1B,2016ATel.9507....1W}\\
SN1000+0216      & 10:00:05.86 & $+$02 16 23.6 & SLSN-II & 3.899 & \cite{2012Natur.491..228C}\\
SN2013am         & 11:18:56.94 & $+$13:03:50.0 & LL IIP$\rm^a$& 0.003 &\cite{2013CBET.3440....1N,2018MNRAS.473.3863L}\\
\hline 
\multicolumn{6}{c}{AAD} \\ 
\hline
Gaia16aye        & 19:40:01.10 & $+$30:07:53.4 & ULENS, CV & & \cite{2016ATel.9376....1B,2016ATel.9507....1W}\\
SN2016fbo        & 01:01:35.54 & $+$17:06:04.3 & Ia$\rm^b$ & 0.030 & \cite{2018MNRAS.475..193F}\\
SDSS-II SN 18391 & 02:22:42.43 & $+$00:25:05.0 & ?Unknown/?Star$\rm^c$ & & \cite{Sako2018}\\
LSQ13ccw         & 21:35:51.64 & $-$18:32:52.0 & Ibn & 0.060 & \cite{2015MNRAS.449.1954P}\\
SN2003gs         & 02:27:38.36 & $-$01:09:35.4 & Ia Pec & 0.005 & \cite{2009AJ....138.1584K}\\
SN2006mr         & 03:22:42.84 & $-$37:12:28.5 & Ia-91bg & 0.006 &  \cite{2010AJ....139..519C}\\
SN2007N          & 12:49:01.25 & $-$09:27:10.2 & Ia-91bg & 0.013 &  \cite{2011AJ....142..156S,2013ApJ...773...53F}\\
SN2007ax         & 08:22:43.26 & $+$22:33:16.9 & Ia-91bg & 0.007 & \cite{2011AJ....142..156S,2013ApJ...773...53F}\\
SN2005dm         & 02:18:39.25 & $-$06:54:10.8 & Ia-91bg & 0.017 & \cite{2005ATel..596....1A,2014ApJ...795..142G}\\
SN2002fb         & 01:57:48.90 & $+$36:20:26.3 & Ia-91bg & 0.016  & \cite{2012AJ....143..126B}\\
PS1-12sk         & 08:44:54.86 & $+$42:58:16.9 & Ibn & 0.054  & \cite{2013ApJ...769...39S}\\
  \hline
\multicolumn{6}{l}{$\rm ^a$ LL IIP --- low-luminosity IIP supernovae.} \\
\multicolumn{6}{l}{$\rm ^b$ LC in the Open Supernova Catalog has a bad quality and contains wrong photometrical points that make it looks anomalous.} \\
\multicolumn{6}{l}{$\rm ^c$ Unknown object in~\cite{Sako2018}; host classified as star by SDSS DR15.} \\
\end{longtable}
\twocolumn

\bibliographystyle{aa}
\bibliography{ref}

\appendix

\section{Active Anomaly Detection Algorithm}
\label{ap:AAD}

We give below, a brief description on how the weights are updated in each iteration of the learning loop. For further details see \citet{Das2018}.

The algorithm starts by training a traditional isolation forest \citep{liu2008isolation}, which requires the user to determine a contamination level, $\tau \in [0,1]$, a percentile used to separate normal objects from anomalies. Once the forest is trained, we denote $q_{\tau}$ the anomaly score corresponding to the chosen contamination level. Each \textit{leaf node} in the forest is subsequently assigned a uniform weight, $w_i=1/\sqrt{N_{\rm nodes}}$. Supposing the average number of leaf nodes per tree is $N_{\rm avt}$, the dimension of the weight vector will be equal to the total number of nodes, ${\rm dim}(\pmb{w}) = N_{\rm trees} \times N_{\rm avt} = N_{\rm nodes}$. We also define a vector $\pmb{z}$ for each object in the data set which also has dimension $N_{\rm nodes}$. Considering the entire set of leaf  nodes as a spatial feature space, each element of $\pmb{z}$ marks the final positions occupied by a given object throughout the forest. In this context, for each object, $\pmb{z}$ is a sparse vector with 0 in all elements corresponding to not occupied leaf nodes. We shall denote the anomaly score of the $i-th$ object as $q_i = \pmb{z}_i \cdot \pmb{w}$. 

Given a data set $\pmb{H}$, we call $\pmb{H}_F  \subseteq  \pmb{H}$ the subset of objects that were already analyzed by the expert, $\pmb{H}_A  \subseteq  \pmb{H}_F$ the set of labelled anomalies and $\pmb{H}_N  \subseteq  \pmb{H}_F$ the set of labelled normal objects. Let $y_i \in [{anomaly, normal}]$ be the label given by the expert to the $i-th$ object. Our goal is to learn the weight vector, $\pmb{w}$, which will allow the labelled anomalies to have a score higher than the score  threshold corresponding to the user choice of $\tau$, $\pmb{w}:q_{\pmb{H}_A} \geq q_{\tau}$.

Using a hinge loss defined as:

\begin{align}
l(q, \pmb{w}; z_i, y_i) = \qquad \qquad  \qquad \qquad \qquad \qquad \qquad  \qquad \qquad \qquad & \nonumber \\
\left\{ 
\begin{array}{l@{}l}
    0 \quad {\rm if} & \pmb{w}\cdot\pmb{z}_i \geq q \quad {\rm and} \quad y_i = {\rm anomaly} \\
    0 \quad {\rm if} & \pmb{w}\cdot\pmb{z}_i < q \quad {\rm and} \quad y_i = {\rm normal} \\
    q - \pmb{w}\cdot\pmb{z}_i \quad {\rm if} \quad& \pmb{w}\cdot\pmb{z}_i < q \quad {\rm and} \quad y_i = {\rm anomaly} \\
    \pmb{w}\cdot\pmb{z}_i - q \quad {\rm if} & \pmb{w}\cdot\pmb{z}_i \geq q \quad {\rm and} \quad y_i = {\rm normal} 
  \end{array} 
\right., & \nonumber 
\\
\end{align}

\noindent the weights for each $t$ iteration of the active learning loop can be found by solving

\begin{eqnarray}
\pmb{w}^{(t)} & = & {\rm arg min}_{\pmb{w}}\left\{ \frac{1}{|\pmb{H}_A|}\sum_{\pmb{z}_i \in \pmb{H}_A}l_{\pmb{q}_{\tau}^{(t-1)}}^i + \frac{1}{|\pmb{H}_N|}\sum_{\pmb{z}_i \in \pmb{H}_N}l_{\pmb{q}_{\tau}^{(t-1)}}^i\right. \nonumber \\
 & + &  \frac{1}{|\pmb{H}_A|}\sum_{\pmb{z}_i \in \pmb{H}_A} l_{\pmb{z}_{\tau}^{(t-1)}}^i + \frac{1}{|\pmb{H}_N|}\sum_{\pmb{z}_i \in \pmb{H}_N} l_{\pmb{z}_{\tau}^{(t-1)}}^i \nonumber \\
 & + & \left. ||\pmb{w} - \pmb{w}_p||^2 \right \}, \label{eq:weights}
\end{eqnarray}

\noindent where
\begin{eqnarray}
l_{\pmb{q}_{\tau}^{(t-1)}}^i & \equiv & l(\hat{q}_{\tau}(\pmb{w}^{(t-1)}), \pmb{w}; \pmb{z}_i, y_i), \\
l_{\pmb{z}_{\tau}^{(t-1)}}^i & \equiv & l(\pmb{z}_{\tau}^{(t-1)}\cdot\pmb{w}, \pmb{w};\pmb{z}_i, y_i), \\
\pmb{w}_p & \equiv & \left[\frac{1}{\sqrt{N_{\rm nodes}}}, ..., \frac{1}{\sqrt{N_{\rm nodes}}}\right]^T,
\end{eqnarray}

\noindent $\pmb{z}_{\tau}^{(t-1)}$ marks the final leaf position for the object at the quantile anomaly score threshold for iteration $t-1$ and $\hat{q}_{\tau}(\pmb{w}^{(t-1)})$ denotes its anomaly score\footnote{By definition, both quantities were calculated with $\pmb{w}=\pmb{w}^{(t-1)}$.}.  Equation \ref{eq:weights} was solved using a RMSProp algorithm, a linear loss function and its corresponding gradient\footnote{\url{https://github.com/shubhomoydas/ad_examples/blob/master/ad_examples/aad/aad_loss.py}}.


\section{Visualization of selected anomalies}
\label{ap:figs}
\begin{figure*}
\begin{center}
\includegraphics[scale=0.48]{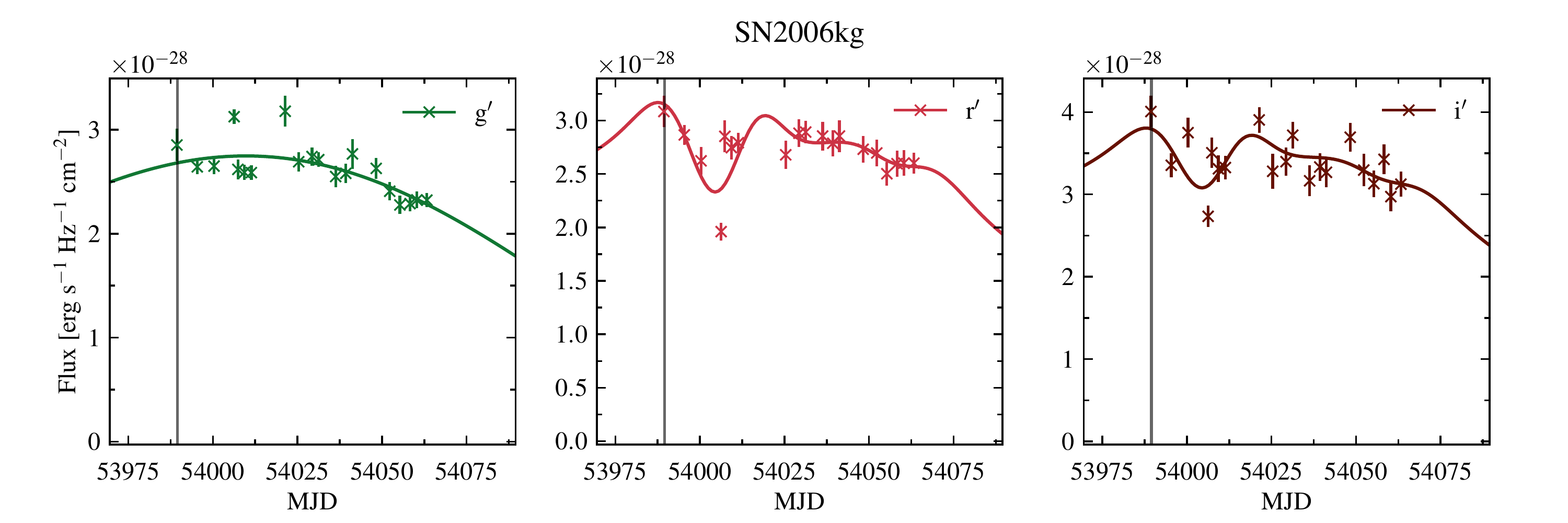}
\caption{Light curves in $g'r'i'$ filters of active galactic nucleus SN2006kg~\citep{Sako2018}. Solid lines are the results of our approximation by \textsc{Multivariate Gaussian Process}. The vertical line denotes the moment of maximum in $r'$ filter.}
\label{SN2006kg}
\end{center}
\end{figure*}

\begin{figure*}
\begin{center}
\includegraphics[scale=0.48]{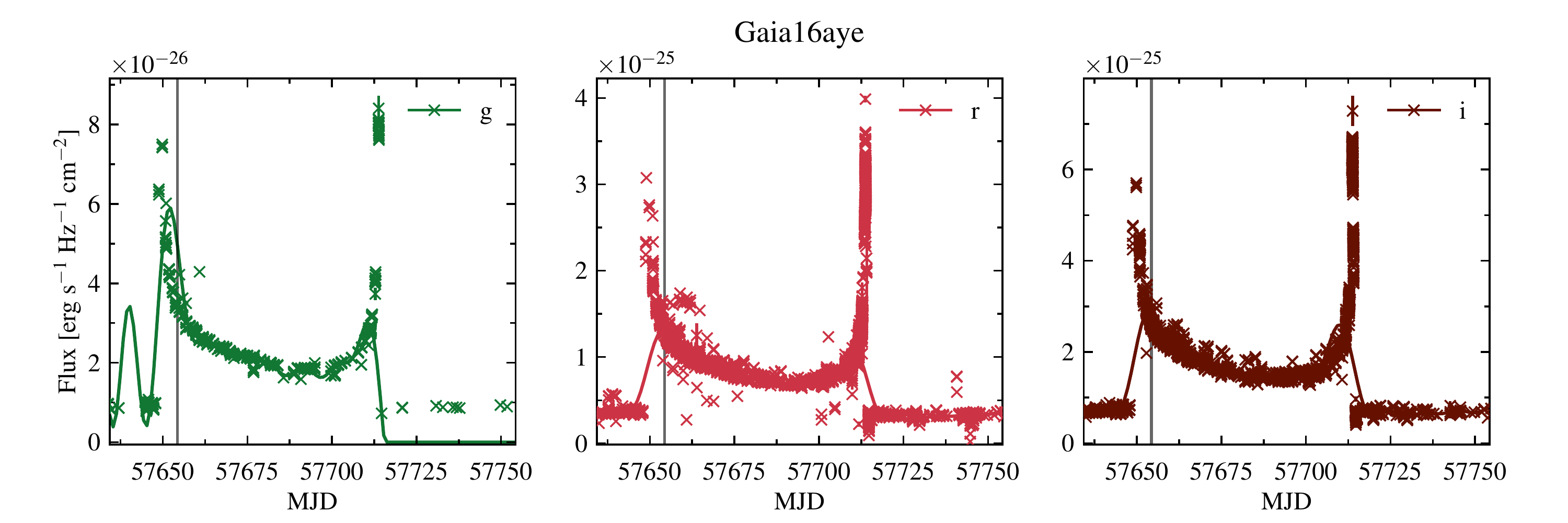}
\caption{Light curves in $gri$ filters of binary microlensing event Gaia16aye (\protect\url{http://gsaweb.ast.cam.ac.uk/alerts/alert/Gaia16aye/followup}). Solid lines are the results of our approximation by \textsc{Multivariate Gaussian Process}. The vertical line denotes the moment of maximum in $r$ filter.}
\label{Gaia16aye}
\end{center}
\end{figure*}

\begin{figure*}
\begin{center}
\includegraphics[scale=0.48]{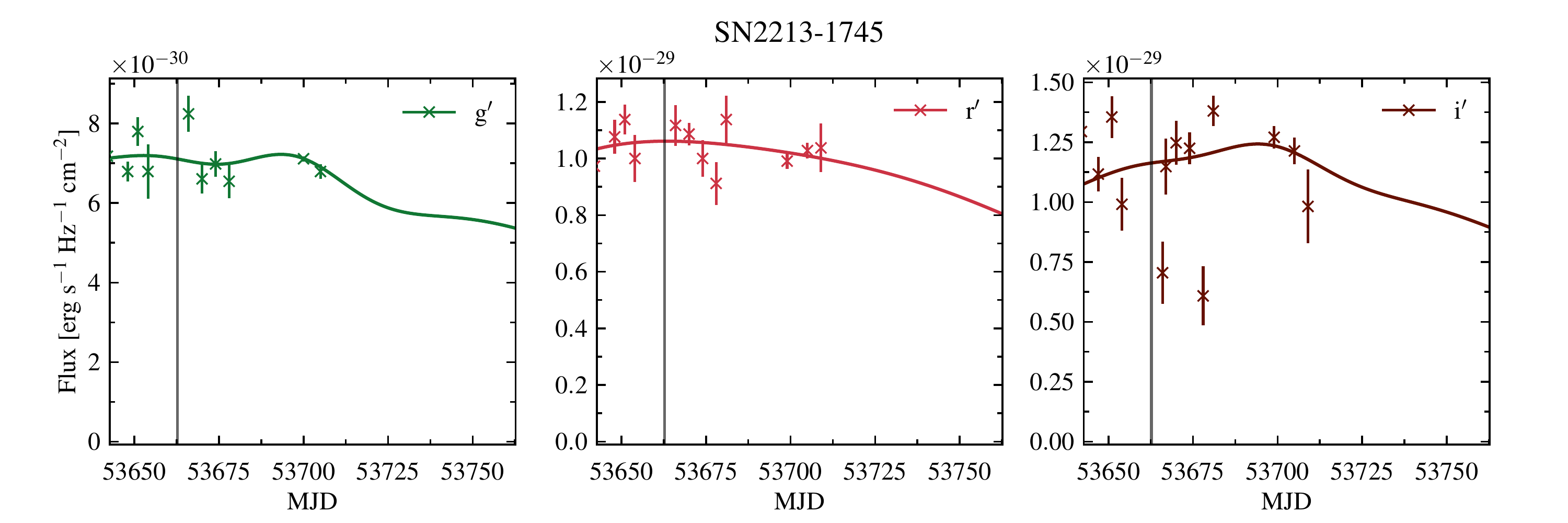}
\caption{Light curves in $g'r'i'$ filters of super-luminous supernova 2213-1745~\citep{2012Natur.491..228C}. Solid lines are the results of our approximation by \textsc{Multivariate Gaussian Process}. The vertical line denotes the moment of maximum in $r'$ filter.}
\label{SN2213}
\end{center}
\end{figure*}

For illustrative purposes, here we show the light curves of three identified anomalies which are potentially interesting for the observer. Two of them, SN 2006kg (Figure \ref{SN2006kg}) and Gaia16aye (Figure \ref{Gaia16aye}) are cases of mis-classification, from which the Open Supernova Catalog partly suffers. The third one, SN2213-1745 (Figure \ref{SN2213}), is an example of a super-luminous supernova --- the rare class of supernovae which has unexplained huge luminosities~\citep{2018SSRv..214...59M}.

\end{document}